\documentclass[referee]{raa}            

\usepackage{caption,subcaption}

\usepackage{graphicx,times}          
\usepackage{natbib}
\usepackage{overpic}
\usepackage{ulem} 
\graphicspath{{./}{figures/}}
\usepackage{amssymb,amsmath}
\bibpunct{(}{)}{;}{a}{}{,}
\usepackage{hyperref}
\hypersetup{colorlinks,				
	linkcolor=blue,
	citecolor=blue}

\begin{document}

   \title{{\bf Quantifying} the anisotropic density structure of the Central Molecular Zone -- a 2D correlation function approach
}

   \volnopage{Vol.0 (20xx) No.0, 000--000}      
   \setcounter{page}{1}          

   \author{Xiu-Yu Cai
      \inst{1,2}
   \and  Guang-Xing Li
      \inst{3}
   \and Lei Qian
      \inst{1,2}
   }

   \institute{University of Chinese Academy of Sciences, Beijing 100049, China\\
   \and
   National Astronomical Observatories, Chinese Academy of Sciences,
             Beijing 100101, China; Lei Qian: {\it  lqian@nao.cas.cn}\\
        \and
      South-Western Institute for Astronomy Research, Yunnan University
        Kunming, 650500 Yunnan, P.R. China; {\it gxli@ynu.edu.cn}\\
\vs\no
   {\small Received~~20xx month day; accepted~~20xx~~month day}}

\abstract{The Central Molecular Zone (CMZ) is a ring-like structure sitting at the center of the Milky Way. Using the 870 $\mu$m continuum map from the APEX Telescope Large Area Survey of the Galaxy (ATLASGAL), 
	we study the anisotropy of the density structure of the gas in the Central Molecular Zone (CMZ) using the 2D correlation function. To quantify the spatial anisotropy, we define the critical angle $\theta_{\rm half}$, as well as the anisotropy parameter
$A\equiv \frac{\pi}{4\theta_{\rm half}}-1$. We find that the density structure is strongly anisotropic at the large scale ($\sim$ 100 pc), and the degree of spatial anisotropy decreases with the decreasing scale. At the scale of $\sim$ 10 pc, the structure is still mildly anisotropic. In our  analyses, we provide a quantitative description of the anisotropic density structure of gas in the CMZ, and the formalism can be applied to different regions to study their differences.
}

   \authorrunning{Xiu-Yu Cai, Guang-Xing Li\& Lei Qian }            
   \titlerunning{Quantifying the anisotropic density structure of the Central Molecular Zone}  

   \maketitle

%
%
\section{Introduction}           
\label{sect:intro}
Situated at the center of the Milky Way, the Central Molecular Zone (CMZ) is an unusually dense molecular cloud complex with a size of a few hundred parsecs  \citep{2002A&A...384..112L}.
A total mass of  $3\times 10^{7} M_{\odot}$ is found inside this region \citep{1998A&A...331..959D, 2014prpl.conf..125M}.
Observations indicate that the gas in the CMZ has high volume densities  (with a mean value of $\sim 10^4 {\ \rm cm^{-3}}$,  \citealt{2020ApJ...897...89L}) and high column densities (about $\sim 10^{23} {\ \rm cm^{-2}}$, \citealt{1994ApJ...424..189L}). 

In spite of the widespread presence of dense gas, the star formation efficiency (SFE) is about $10-100$ times lower than the standard values \citep{2013ApJ...765L..35K, 2013MNRAS.429..987L,2015MNRAS.446.2468E,2018MNRAS.478.3380J}. The dynamics in this region can be affected by a  variety of processes, such as gravitational instability, turbulence, tidal force, cloud-cloud collision and shear etc. \citep{2013MNRAS.429..987L,2018MNRAS.478.3380J, 2019MNRAS.484.5734K}. 
 
The shear on a cloud will cause a velocity difference between its near side and its far side to the Galactic center, which stretches the gas into long streams. 
The strength of shear can be quantified using the  shear timescale, which is $t_{\rm shear} = (\partial \Omega / \partial r \times r)^{-1}$. In the Milky way, shear is believed to be responsible for creating large-scale filamentary structures \citep{2006MNRAS.367..873D}.
 Some first hints on the importance of shear came from the discovery of kpc-sized filamentary structures \citep{2013A&A...559A..34L,2014ApJ...797...53G,2015MNRAS.450.4043W}.
Further observations have found that the filamentary structures of sizes of a few to a few tens of pc tend to stay parallel to the Galactic disk \citep{2016A&A...591A...5L,2015MNRAS.450.4043W}, indicating that shear is dynamically important on these scales. In some cases, shear can play a dominant role in determining the star formation activity: recent results from \citet{2020ApJ...897...89L}
indicate that shear alone is responsible for the observed low level of star formation seen in the CMZ region.

One way to reveal the role of shear is to study the alignment of filamentary structures \citep{2016A&A...591A...5L,2015MNRAS.450.4043W}. However, this approach, although effective, is cumbersome to implements. Besides, the evolution of the interstellar medium is a multi-scale process, and ideally, we would like to know the role of shear over a range of scales. As studying the role of shear using the alignment of filamentary structures only allows us to probe the scales comparable to the lengths of the filaments, better methods are needed. 
  In this paper, we develop a formalism to quantify the anisotropy of the density structure of the CMZ quantitatively using the 2D correlation function, with which the role of shear can be studied over a range of scales. In section 2, we present the data. In section 3, we describe the method and present the results. In section 4, we give a conclusion. 

\begin{figure} 
 \centering
 \begin{subfigure}[t]{3in}
  \centering
  \includegraphics[width=3in]{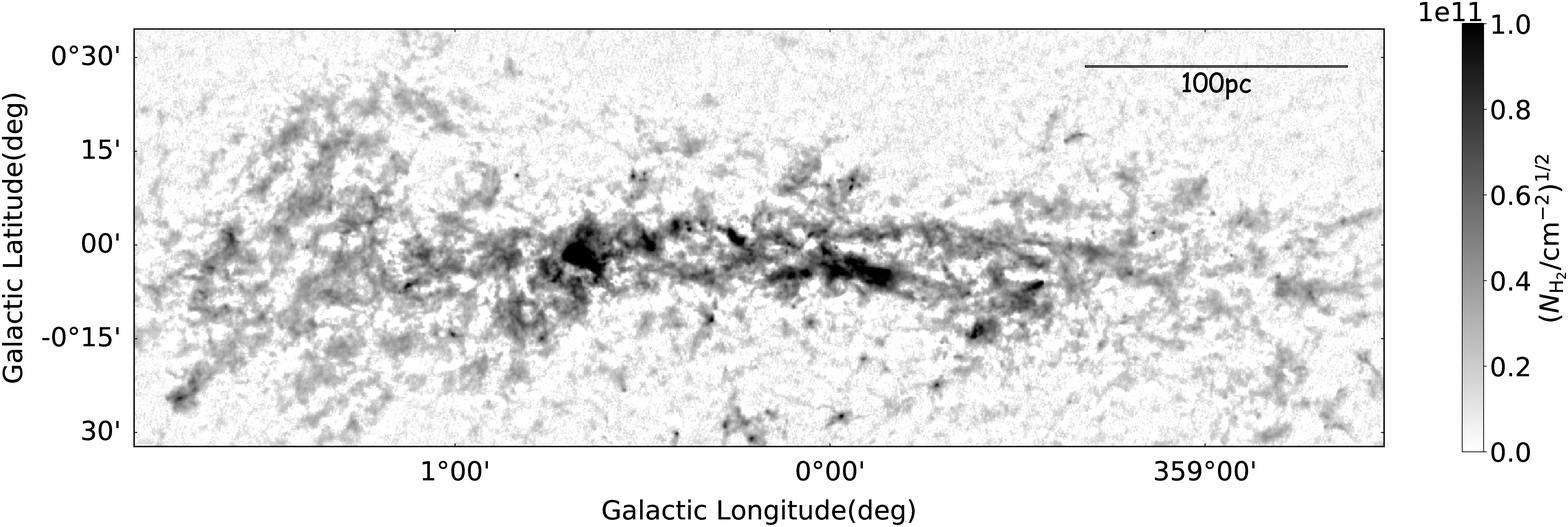}
  \caption{column density}\label{fig:1(a)}
 \end{subfigure}
 \begin{subfigure}[t]{2in}
  \centering
  \includegraphics[width=3in]{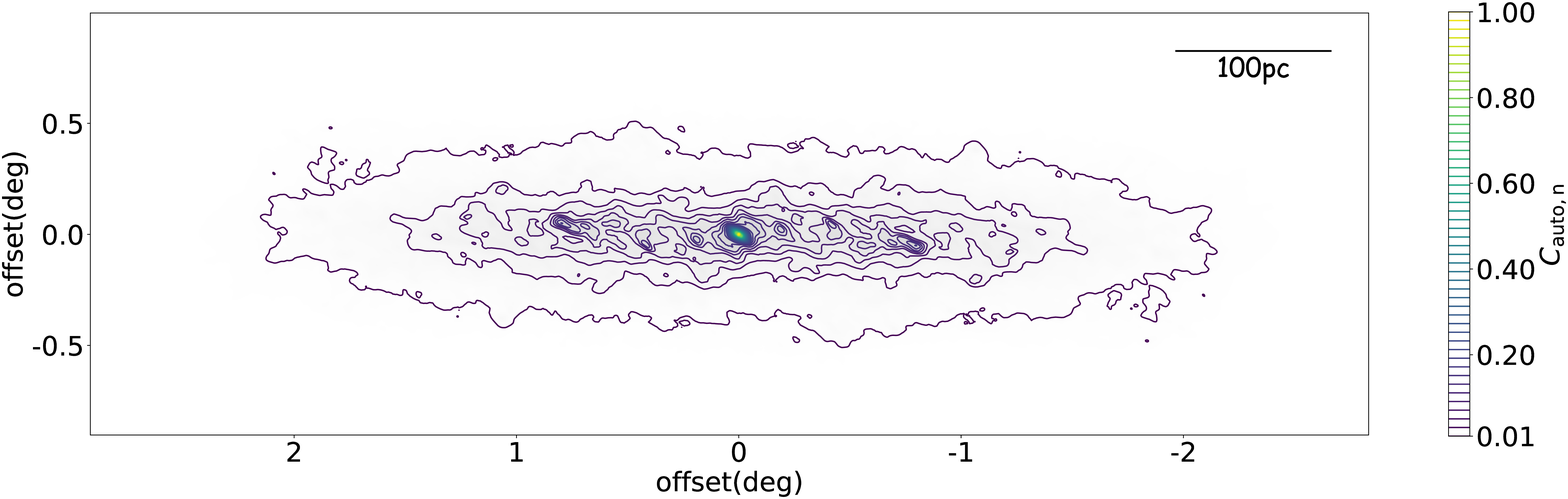}
  \subcaption{anisotropy}\label{fig:1(b)}
 \end{subfigure}
 
 \quad
 \begin{subfigure}[t]{3in}
  \centering
  \includegraphics[width=3in]{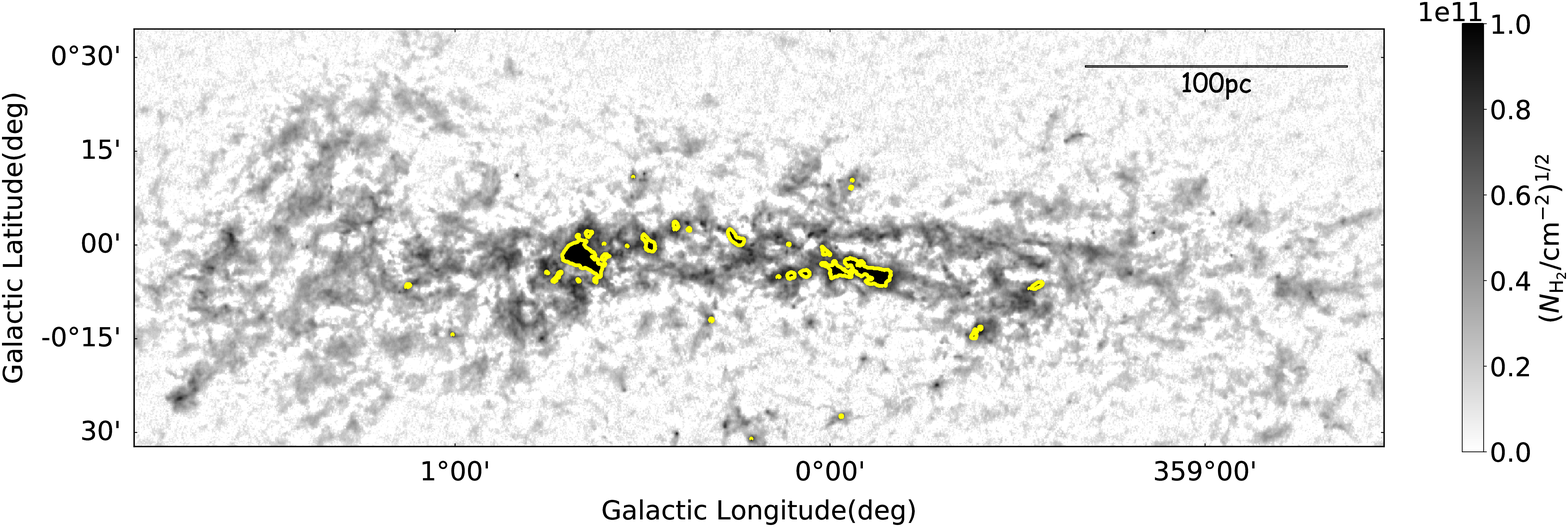}
  \caption{column density\_3}\label{fig:1(c)}
 \end{subfigure}
 \begin{subfigure}[t]{2in}
  \centering
  \includegraphics[width=3in]{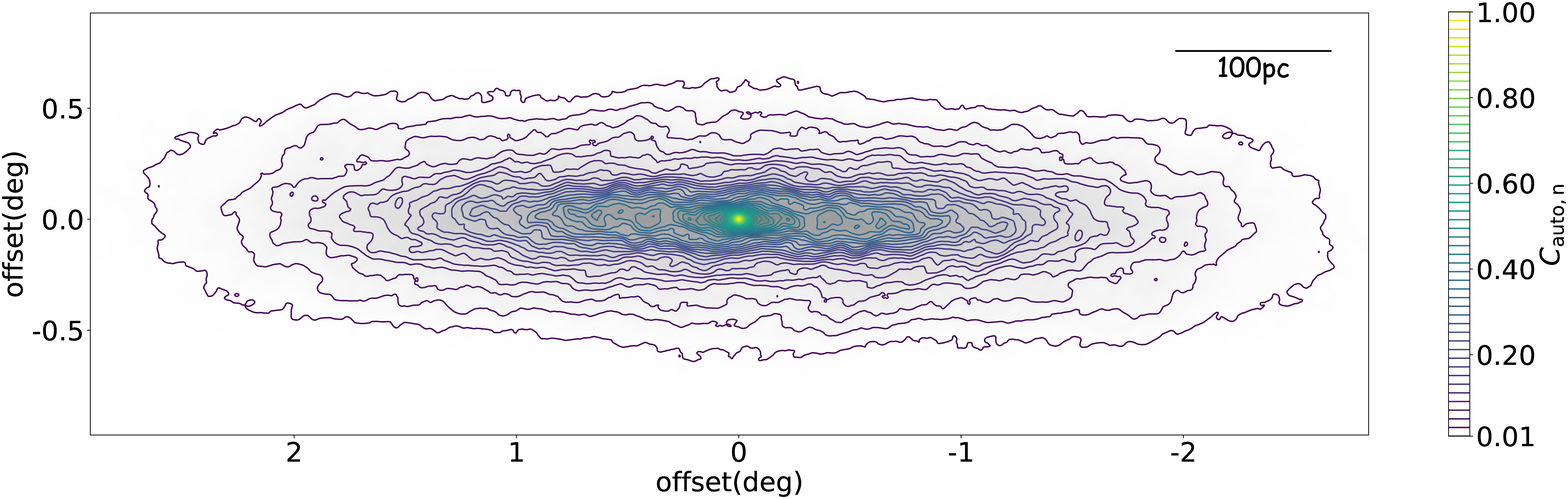}
  \subcaption{anisotropy\_3}\label{fig:1(d)}
 \end{subfigure}
 
 \quad
 \begin{subfigure}[t]{3in}
  \centering
  \includegraphics[width=3in]{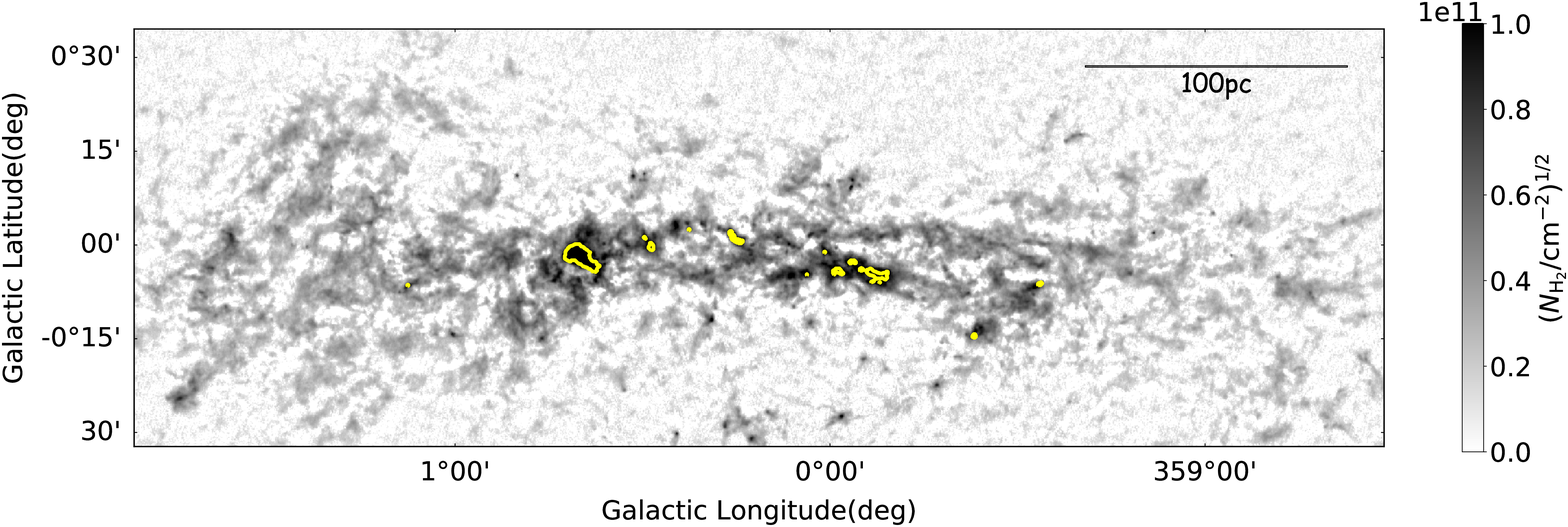}
  \caption{column density\_5}\label{fig:1(e)}
 \end{subfigure}
 \begin{subfigure}[t]{2in}
  \centering
  \includegraphics[width=3in]{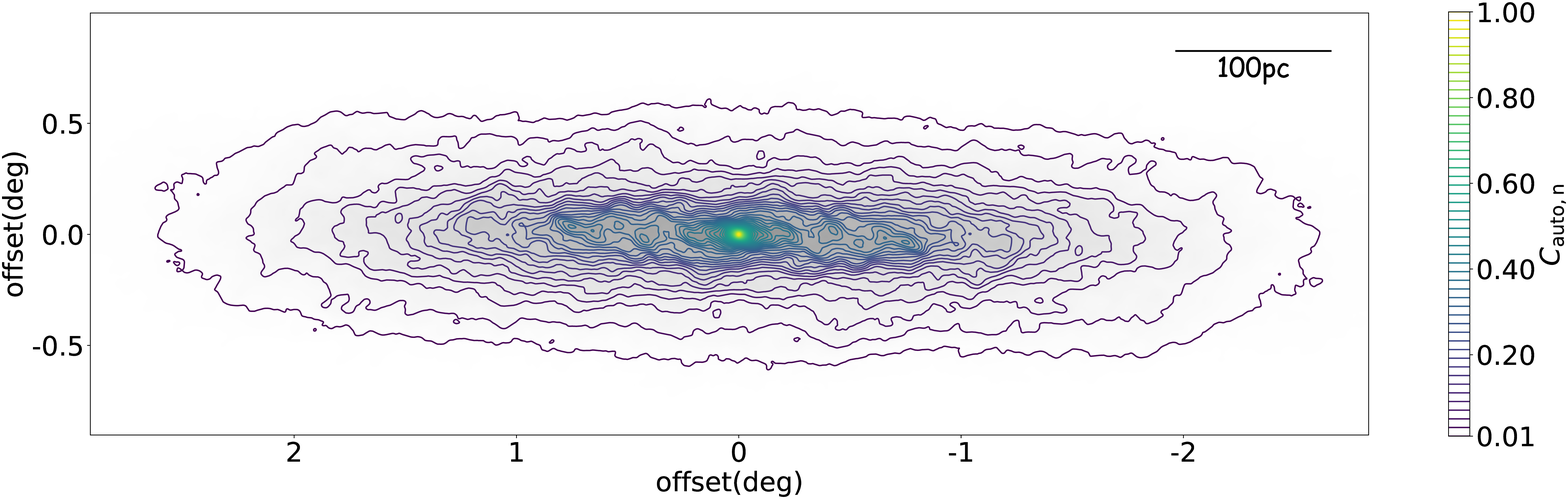}
  \subcaption{anisotropy\_5}\label{fig:1(f)}
 \end{subfigure}
 \caption{{\bf Left panels:} Column density distribution in the Central
Molecular Zone of the Galactic center observed by the survey:the APEX Telescope
Large Area Survey of the Galaxy (ATLASGAL) at $870 {\ \mu\rm m}$, with a beam size of
$19.2\arcsec$ and a typical noise level of 50-70 $\rm mJy/beam$. The
horizontal size bar in upper right corner indicates a length of 100 pc. The yellow contours
in Figure \ref{fig:1(c)} show the region where the flux intensity is more than 3 $\rm Jy/beam$
($8.31\times10^{21}{\ \rm{cm^{-2}}}$). The yellow contours in  Figure \ref{fig:1(e)} shows the
region where the flux intensity is more than 5 $\rm Jy/beam$ ($1.385\times10^{22}{\
\rm{cm^{-2}}}$).
 {\bf Right panels:} 2D correlation functions $C_{\rm auto,\rm n}$. In Figure \ref{fig:1(b)}, 
the correlation function is computed using the emission map. In Figures \ref{fig:1(d)} and Figure \ref{fig:1(f)}, we plot the correlation function computed from clipped emission maps where we set the value of regions where  I $>$ 3, 5 $\rm Jy/beam$ to $I_{\rm max}$ = 3, 5 $\rm Jy/beam$, which correspond to
 $N_{\rm H_2} = 8.31\times10^{21}{\ \rm{cm^{-2}}} $ and $ 1.385\times10^{22}{\ \rm{cm^{-2}}}$.}
\end{figure}

\section{DATA}
\label{sect:Obs}
We use the $870 {\ \mu\rm m}$ map from the APEX Telescope Large Area Survey of
the Galaxy (ATLASGAL) \citep{2009A&A...504..415S}. The observations were carried
out with the APEX 12 m submillimeter telescope in dust emission continuum with
an angular resolution of $19.2\arcsec$ and a sensitivity of 50 $\rm Jy/beam$. We
assume that  the CMZ region has a mean distance of 8.2 kpc, estimated from
the updated distance to the Sgr A from \citet{2019A&A...625L..10G}. The
corresponding spatial resolution is $\sim$ 0.76 pc. The size of the
selected region is about 477 pc $\times$ 159 pc. The maps contain
contaminations from the fore/background, and this amounts to around 10 \%  of the total flux \citep{2020ApJ...897...89L}. Thus, the contributions from fore/background emission to the overall correlation should be insignificant.

\section{METHODS AND RESULTS}
\label{sect:data}
\subsection{The correlation function}

Our data is taken from \citet{2009A&A...504..415S}, from which the mass surface density can be calculated by using 
\begin{equation}
N_{\rm H_2}=\frac{F_{\nu} R_0}{B_{\nu}(T_{\rm D})\Omega \kappa \mu m_{\rm H}},
\end{equation}
where ${F_{\nu}}$ is the flux density, $R_0\sim 100$ is the gas-to-dust ratio, $\Omega$ is the solid angle of the telescope beam, $\mu \sim 2.8$ is the mean molecular weight of the interstellar medium with respect to hydrogen molecules, \citep{2008A&A...487..993K}, and ${m_{H}}$ is the mass of an hydrogen atom. We adopt a uniform temperature of {\bf $T_{\rm D}=20 {\ \rm K}$} \citep{2016A&A...586A..50G,2010A&A...518L.100M,2011MNRAS.416.2932T}.   
At 870 $\mu$m, $\kappa = 1.85 {\ \rm cm^{2}\ g^{-1}}$ \citep{1994A&A...291..943O} and

\begin{equation}
B_{\nu}(T_D)={\frac{2h{\nu}^3}{c^2}{\frac{1}{{e^{\frac{h{\nu}}{kT}}}-1}}}\;.
\end{equation}
Here, we have assumed a uniform temperature for the whole region. In reality, the temperature varies by around 5 K, leading to an uncertainty of about 20 \% for individual regions. As the errors of surface density caused by temperature variations are small compared to the intensity variations, the contribution from temperature to the overall correlation should be minimal.

To quantify the anisotropy of the density structure, we evaluate 
$C_{\rm auto}$, which is the  2D correlation  function. Assuming that $f(x,y)$ represents the intensity distribution in the $x$--$y$ plane, its Fourier transform  is
\begin{equation}
f(k_{1},k_{2})=\int_{-\infty}^{\infty} \int_{-\infty}^{\infty}f(x,y)e^{-i(k_{1}x+k_{2}y)}\,dxdy.
\end{equation}
We first calculated the power spectrum in the $k$-space. $f^\dagger$ is the conjugation of $f$.
\begin{equation}
P(k_{1},k_{2})=f(k_{1},k_{2}) {f^\dagger(k_{1},k_{2})}.
\end{equation}
Then the correlation function in the real space, $C_{\rm auto}$ is obtained by the inverse transform
\begin{equation}
C_{\rm auto}(x,y)=\frac{1}{4\pi^2}\int_a^b \int_a^b P(k_{1},k_{2})e^{i(k_{1}x+k_{2}y)}\,dk_{1}dk_{2}.
\end{equation}

The column density map of the CMZ is shown in Figure \ref{fig:1(a)}.
The normalized 2-D correlation function ( $C_{\rm auto,\rm n}\equiv C_{\rm auto}/C_{\rm auto,\rm max}$) based on the flux data is shown in Figure \ref{fig:1(b)}. Note that calculations done in the Fourier space assumes a periodic boundary condition. To minimize boundary effects, we add zero paddings around our maps before performing the calculations. We note that in some cases, a very significant 
 amount of emission is contained in a very small region (e.g. the Sgr B2 region). To evaluate their contributions to the overall correlation, we experiment with performing clipping operations to the data at regions where the flux is larger than a certain threshold  of 3 $\rm Jy/beam$ ($8.31\times10^{21}{\ \rm{cm^{-2}}}$) and 5 $\rm Jy/beam$ ($1.385\times10^{22}{\ \rm{cm^{-2}}}$), separately. This is achieved by setting the values of regions where the flux is above a threshold to that threshold. We set flux data more than 3, 5 $\rm Jy/beam$ ($8.31\times10^{21}{\ \rm{cm^{-2}}}$, $1.385\times10^{22}{\ \rm{cm^{-2}}}$) to $I_{\rm max}$ = 3, 5 $\rm Jy/beam$ ($8.31\times10^{21}{\ \rm{cm^{-2}}}$, $1.385\times10^{22}{\ \rm{cm^{-2}}}$). The regions where the flux data more than 3, 5 $\rm Jy/beam$ ($8.31\times10^{21}{\ \rm{cm^{-2}}}$, $1.385\times10^{22}{\ \rm{cm^{-2}}}$) are marked yellow in Figure \ref{fig:1(c)} and Figure \ref{fig:1(e)}, separately. The clipping operation effectively reduces the dynamic range of the maps. The correlation functions computed from these clipped maps are presented in Figure \ref{fig:1(d)} and Figure \ref{fig:1(f)}. By comparing the clipped results ( Figure \ref{fig:1(d)} and
 Figure \ref{fig:1(f)} ) to the unclipped ones ( Figure \ref{fig:1(b)}), we are able to evaluate the contribution to the correlation function from regions of different surface densities.

From all these correlation functions, we observe that the contours in the centre (at small scales) are nearly roundish, but at larger scales, the contours are elliptical where the long axes of the ellipses are aligned with the mid-plane of the Milky Way. This is an indication that shear is dynamically important in the region. 

\subsection{Quantifying the spatial anisotropy}

\begin{figure}[h]
  \begin{minipage}[t]{0.55\linewidth}
   \includegraphics[width=79mm]{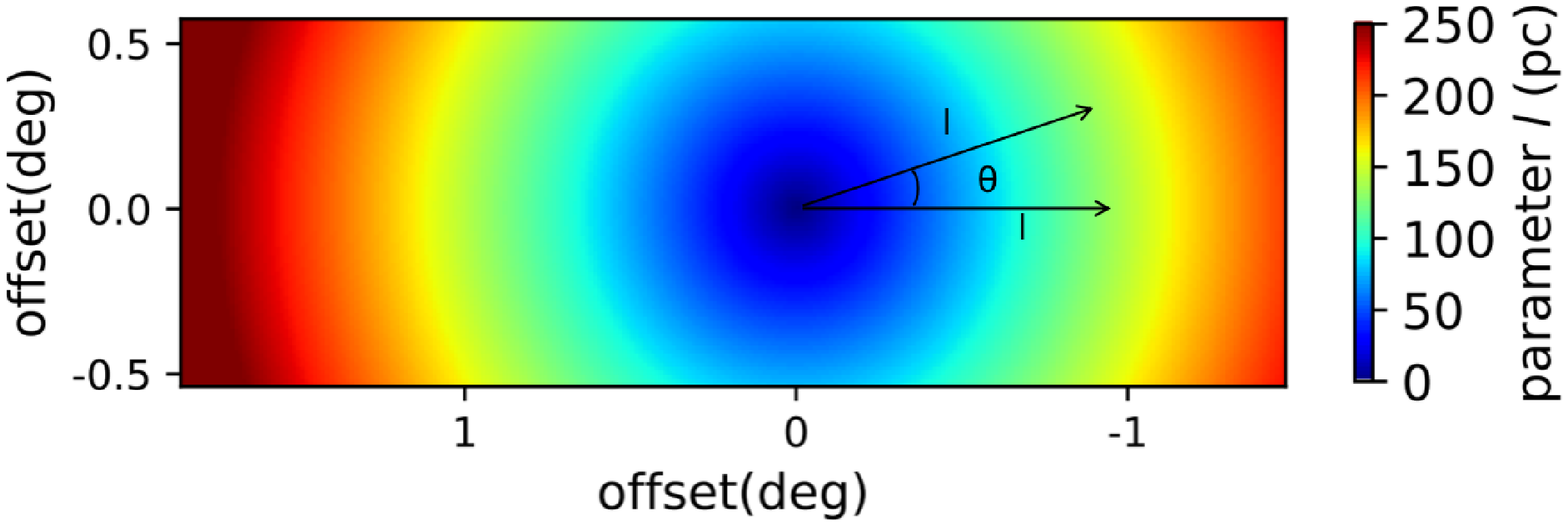}
   \subcaption{}
   \label{fig 2a}
  \end{minipage}%
  \begin{minipage}[t]{0.45\textwidth}
  \centering
   \includegraphics[width=79mm]{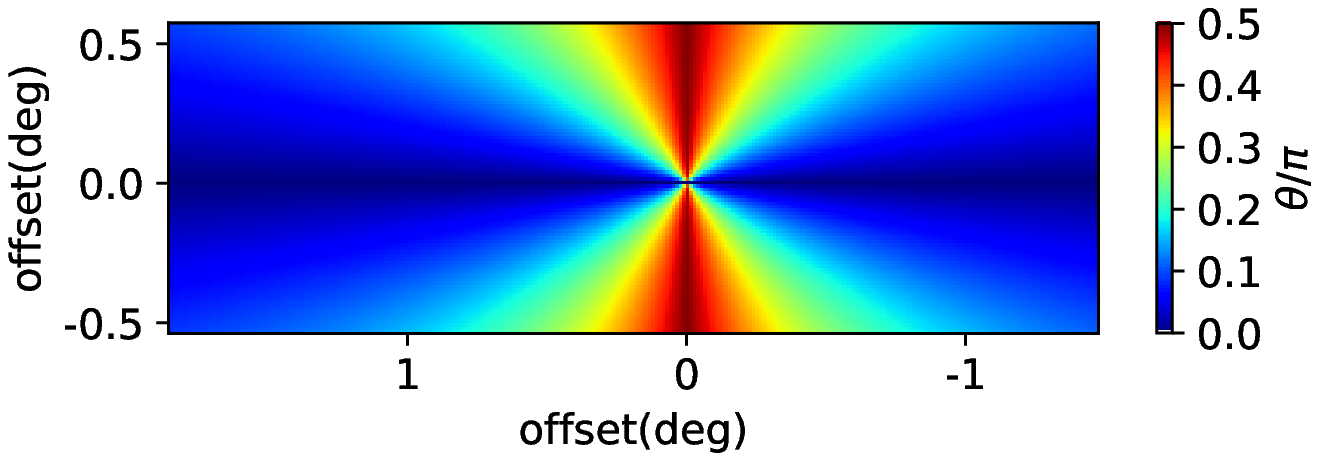}
   \subcaption{}
   \label{fig 2b}
     \end{minipage}%
\caption{(a) The region is divided with different radii and $l$ is the distance from the origin. (b) The region is divided with different angles, we symmetrize it into the range of 0 to $\frac{\pi}{2}$, the colorbar is $\theta/\pi$.}
\end{figure}

To further quantify the spatial anisotropy measured as a function of the scale, we divide the region into rings of  different
radii. Each ring is characterized by the  $l$ parameter, which is the distance from the origin ( Figure \ref{fig 2a}). For each ring, we plot the value of the correlation function against $\theta$, which is the angle  measured with respect to the Galactic mid-plane.  
We have considered all the points. And $\theta$ has been transformed to the range (0,$\frac{\pi}{2}$) (Figure \ref{fig 2b}). Due to symmetry, the $\theta$ is between 0 and $\pi$. Additional, we assume that the structure of the CMZ is symmetric had we turned it upside down (horizontal mirror symmetry).
Thus, we only need to plot $\theta$ between 0 and $\frac{\pi}{2}$.

\begin{figure}[h]
  \begin{minipage}[t]{0.55\linewidth}
   \includegraphics[width=80mm]{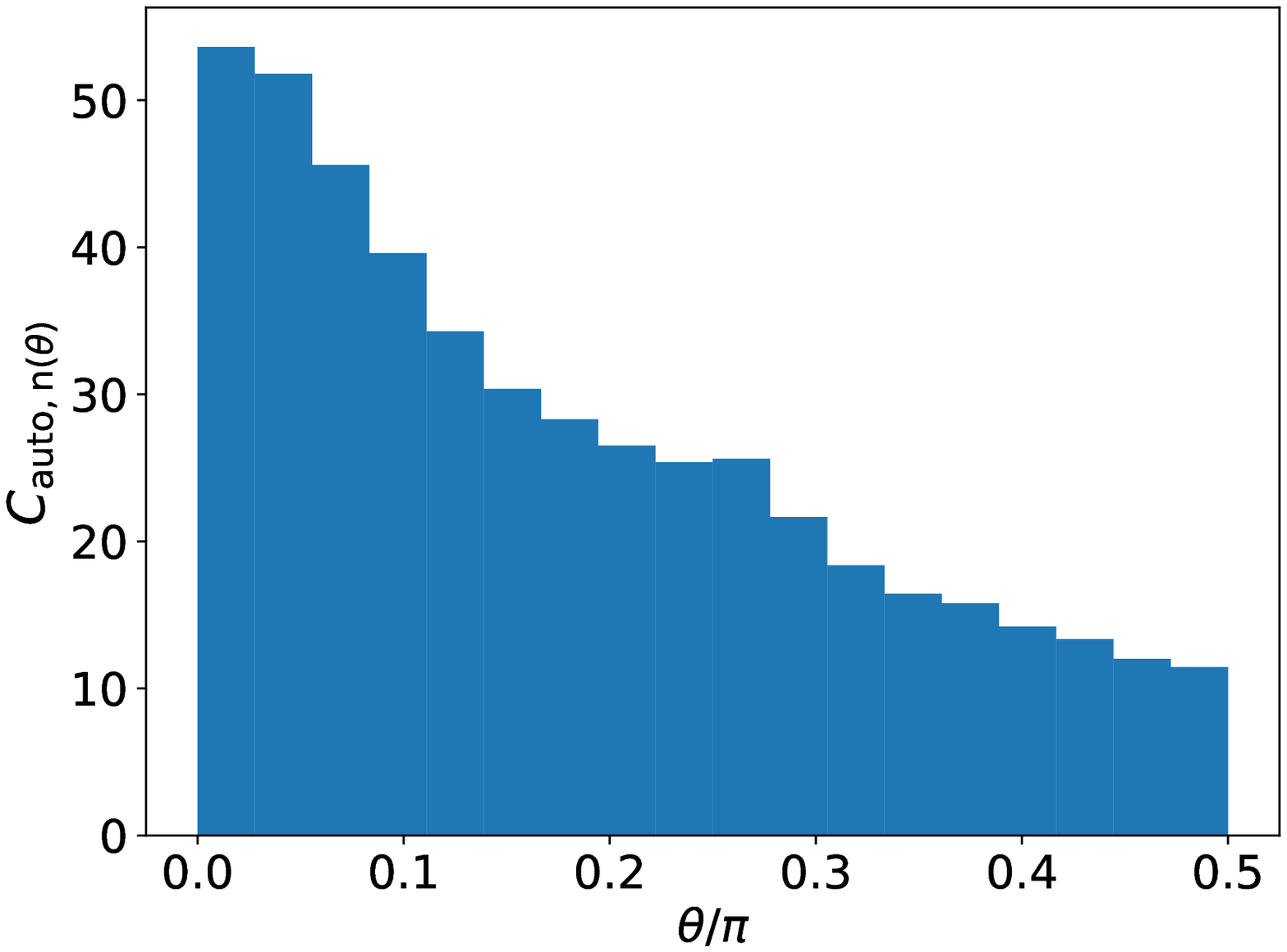}
   \subcaption{}
   \label{fig:2(a)}
  \end{minipage}%
  \begin{minipage}[t]{0.45\textwidth}
  \centering
   \includegraphics[width=80mm]{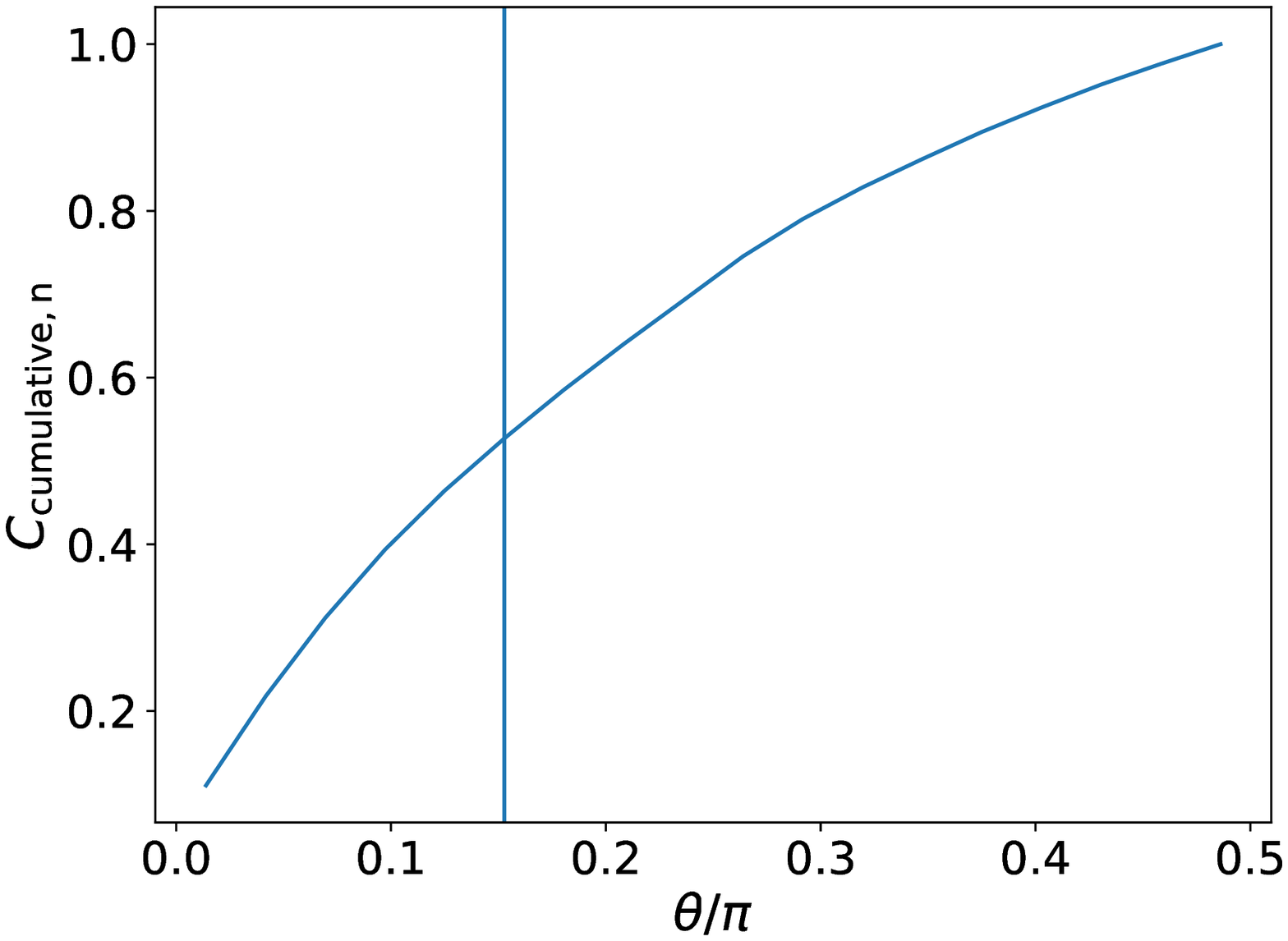}
   \subcaption{}
   \label{fig:2(b)}
     \end{minipage}%
\caption{(a) The distribution of $C_{\rm auto,\rm n}$ against $\theta$. (b)  $C_{\rm cumulative,\rm n}$- $\theta/\pi$ relation. $C_{\rm cumulative,\rm n}$  is the normalized cumulative correlation function. The vertical line shows where the $C_{\rm cumulative,\rm n}$ is 0.5.}
\end{figure}

As an example, we plot the results from a ring  at  $l \sim 33.4 - 35.8$ pc. The ring width $\Delta l$ is 10 pixels, about 2.4 pc. {In Figure \ref{fig:2(a)}, we plot  the $C_{\rm auto,\rm n}$ against $\theta$.} Here, the spatial anisotropy  can be seen from the fact that the the correlation is stronger along the $l$ direction where $\theta =0$, but becomes weaker as $\theta$ increases.

\begin{figure}[h]
  \begin{minipage}[t]{0.55\linewidth}
   \includegraphics[width=80mm]{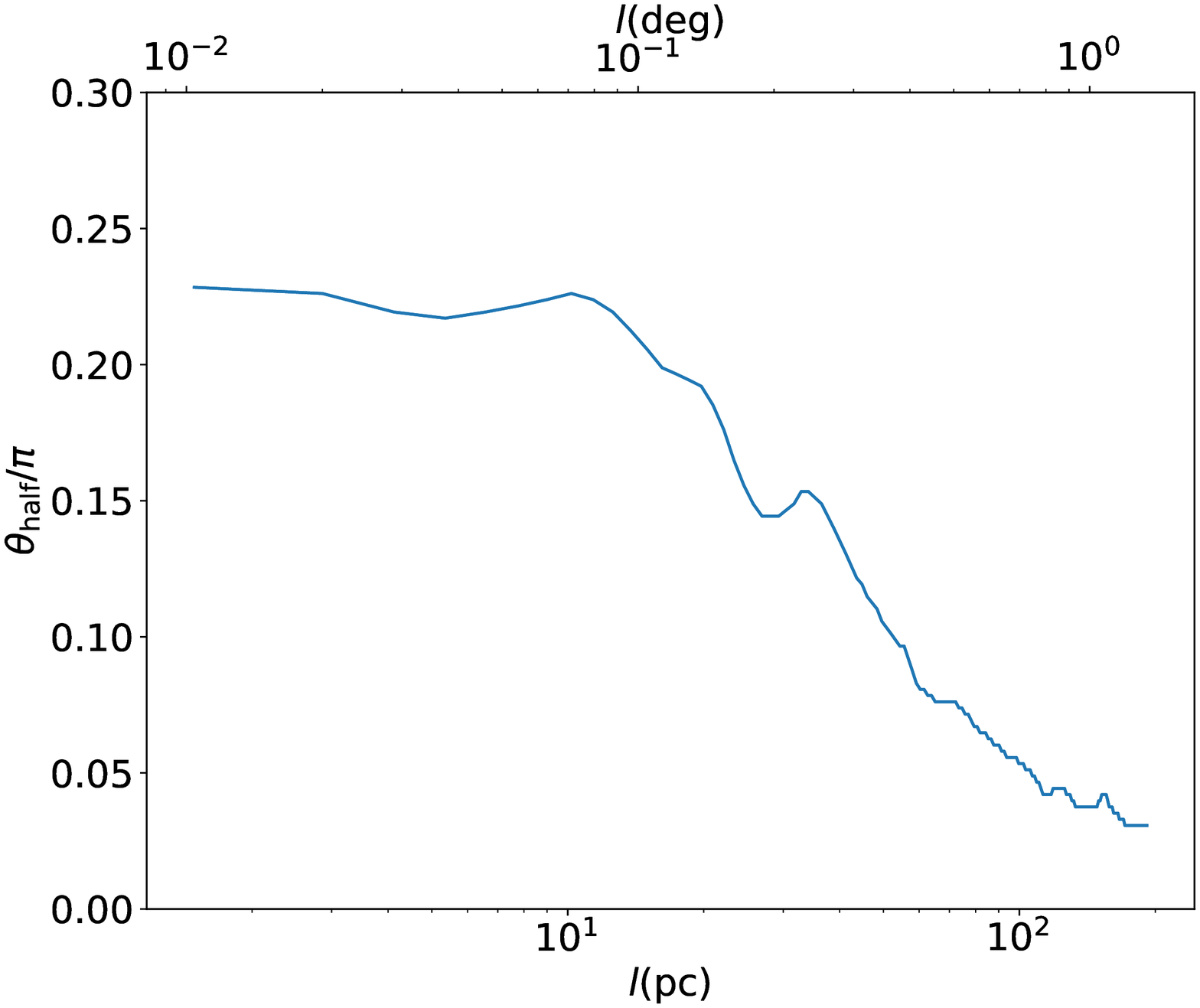}
   \subcaption{}
   \label{fig 3a}
  \end{minipage}%
  \begin{minipage}[t]{0.45\linewidth}
  \centering
   \includegraphics[width=80mm]{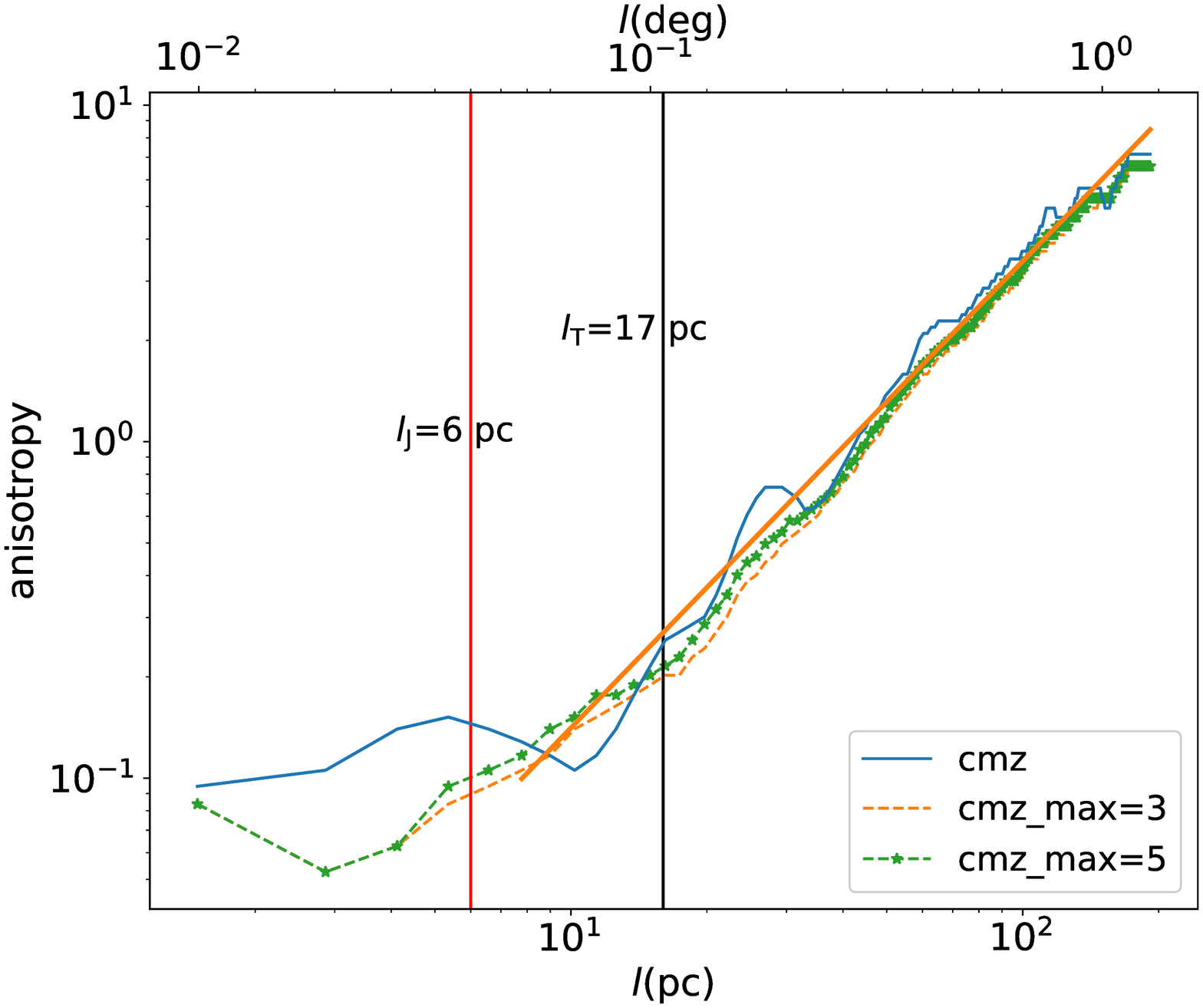}
   \subcaption{}
   \label{fig 3b}
  \end{minipage}%
\caption{(a) $\theta_{\rm half}/\pi$ - $l$ relation. $\theta_{\rm half}$ decreases with $l$  in the CMZ. (b) Anisotropy-$l$ relation. The left vertical line indicates where the Jeans lengths $l_{\rm J}$ (6 pc) and the right vertical line represents the Toomre lengths $l_{\rm T}$ (17 pc) estimated by \citet{2016MNRAS.463L.122H}. The solid line is the anisotropy-$l$ distribution calculated with raw data. The dashed line is the anisotropy-$l$ distribution calculated by setting raw data more than 3 $\rm Jy/beam$ ($8.31\times10^{21}{\ \rm{cm^{-2}}}$) to $I_{\rm max}$ = 3 $\rm Jy/beam$ ($8.31\times10^{21}{\ \rm{cm^{-2}}}$). The dashed line with stars is the anisotropy-$l$ distribution calculated by setting raw data more than 5 $\rm Jy/beam$ ($1.385\times10^{22}{\
\rm{cm^{-2}}}$) to $I_{\rm max}$ = 5 $\rm Jy/beam$ ($1.385\times10^{22}{\
\rm{cm^{-2}}}$). The straight line at scales larger than 10 pc is the linear fit of data, ${\rm log}_{10} \; A \approx 1.4\; {\rm log}_{10} \;l - 2.2$.} 
\end{figure}

To quantify the spatial anisotropy at scale $l$, we define the so-called half correlation angle $\theta_{\rm half}$, which is the critical angle within which half of the correlation function is contained. {To derive  $\theta_{\rm half}$, we define the so-called cumulative correlation function $C_{\rm cumulative} (\theta) = \int C_{\rm auto,\rm n}(\theta) {\rm d} \theta$, then we normalize it, obtaining
the normalized cumulative correlation function $C_{\rm cumulative,\rm n}(\theta) =C_{\rm cumulative}(\theta)/C_{\rm cumulative,\rm max} (\theta)$.
The $\theta_{\rm half}$  is obtained by solving c The procedure is illustrated in Figure \ref{fig:2(b)}. For isotropic structures, $\theta_{\rm half} = \frac{\pi}{4}$, whereas for structures that are preferentially aligned with the disk mid-plane, $\theta_{\rm half} < \frac{\pi}{4}$. An example of how we derive $\theta_{\rm half} $ is presented in Figure \ref{fig:2(b)}.

The above-mentioned exercise allow us to study how the spatial anisotropy evolve as the function of the scale. In Figure \ref{fig 3a}, we plot $\theta_{\rm half}$ against the scale, and where 
$\theta_{\rm half}$ decreases with the increasing scale. This indicates that the spatial anisotropy is stronger at larger scales.

We further define the the anisotropy parameter 
\begin{equation}
A\equiv \frac{\pi}{4\;\theta_{\rm half}}-1\;.
\end{equation}
$A>0$ means that at the scale of interest, the density structure is anisotropic, and the value of $A$ measures the degree of anisotropy. 
For an homogeneous region, the correlation strength is evenly distributed between 0 and $\frac{\pi}{2}$. 
$\theta_{\rm half} = \frac{\pi}{4}$, $A$ will be 0. As $\theta = 0$ corresponds to the direction along which the correlation concentrates, in most cases, $\theta_{\rm half} < \frac{\pi}{4}$ and  $A >0$. 
In the Figure \ref{fig 3b}, we plot $A$ against the scale.  
We also plot regions where we have chosen different $I_{\rm max}$. 
We note that at scales below 10 pc, the anisotropy parameter depends on the value of  $I_{\rm max}$. Thus, we should only interpret results from  scales larger than 10 pc, as only in this range do results from different $I_{\rm max}$ converge. At scales larger than 10 pc, we perform a fit to our data and find  ${\rm log}_{10} \; A \approx 1.4\; {\rm log}_{10} \;l - 2.2$. The anisotropy is strong on the large scale, and it decreases with decreasing scale. At around $l$ = 10 pc,  the density structure is still moderately anisotropic.

We further add the two vertical lines indicating the Toomre length $l_{\rm T}\approx 17$ pc and the Jeans length $l_{\rm J}\approx 6$ pc \citep{2016MNRAS.463L.122H}, respectively. The Jeans length is the length scale above which the gravity can induce collapse, and the Toomre length is the length below which self-gravity is stronger than shear. The density structure is expected to be anisotropic at $l > l_{\rm T}$, and this is confirmed by our results. Apart from this, we can still observe a significant amount of anisotropy at $l_{\rm T} > l >  l_{\rm J}$. 
\section{Conclusions}%
We study the density structure of gas in the Central Molecular Zone by applying the 2D correlation function. We find that the density structure is strongly anisotropic where the correlation is strong along the $l$ direction, suggesting that shear is dynamically important.

To quantify anisotropic density structure, we define the half-correlation angle $\theta_{\rm half}$ and the anisotropy parameter $A\equiv \frac{\pi}{4\theta_{\rm half}}-1\;.$ The density structure is strongly anisotropic at $l = 100\;\rm pc$, and is slightly anisotropic at $l = 10\;\rm pc$. The between 10 pc and 100 pc, we find that  
$${\rm log}_{10}\; A \approx 1.4\; {\rm log}_{10} \, l - 2.2\;.$$

We propose a picture where at the large scale ($l$ $>$ 10 pc), shear is dynamically important such that it can change the density structure of the gas significantly, and its strength diminishes as one moves to  smaller scales. At ($l$ $\lesssim$ 10 pc), as shear should have the  roughly same strength as the self-gravity, thus the gas can enter a state called ``shear-enabled pressure equilibrium''  \citep{2020ApJ...897...89L}.
The formalism developed here can be used to study the role of shear in different regions in a quantitative fashion and reveal the differences.

\begin{acknowledgements}
We thank the anonymous referee for a constructive review report that improved this paper. Lei Qian is supported by NSFC No. U1631237 and
the Youth Innovation Promotion Association of CAS (id. 2018075). Guang-Xing Li is supported by a starting grant from Yunnan University, and NSFC grant W820301904. The paper makes use of data from the ATLASGAL survey carried out by the APEX telescope.
\end{acknowledgements}



\label{lastpage}

\end{document}